\documentclass[conference,twocolumn,10pt]{IEEEtran}
\usepackage{graphicx}
\usepackage{amsmath,amssymb}
\usepackage{dsfont}
\usepackage{cases}
\usepackage{bm}
\usepackage{amsfonts}
\usepackage{times}
\usepackage[font=footnotesize]{caption}
\usepackage{indentfirst}
\usepackage{cite}
\usepackage{caption}
\usepackage{algorithm}
\usepackage{algpseudocode}
\usepackage{booktabs}
\usepackage{subfigure}
\usepackage{multirow}
\usepackage{amsthm}
\usepackage{setspace}

\theoremstyle{remark}
\newtheorem{remark}{Remark}
\theoremstyle{}
\newtheorem{theorem}{Theorem}
\theoremstyle{}
\newtheorem{corollary}{Corollary}

\theoremstyle{}
\newtheorem{definition}{Definition}
\theoremstyle{remark}
\newtheorem{example}{Example}
\theoremstyle{definition}

\makeatletter
\newcommand{\tabcaption}{\def\@captype{table}\caption}
\makeatletter
 \allowdisplaybreaks[4]

\ifCLASSINFOpdf
\else
\fi

\begin{document}

%\title{A New Scheme Attaining Flexible Tradeoff in Distributed Computing}
\title{A Storage-Computation-Communication Tradeoff for Distributed
Computing}

%\author{Michael~Shell,~\IEEEmembership{Member,~IEEE,}
%        John~Doe,~\IEEEmembership{Fellow,~OSA,}
%        and~Jane~Doe,~\IEEEmembership{Life~Fellow,~IEEE}% <-this % stops a space
%\thanks{M. Shell is with the Department
%of Electrical and Computer Engineering, Georgia Institute of Technology, Atlanta,
%\thanks{Manuscript received April 19, 2005; revised December 27, 2012.}}
\author{\IEEEauthorblockN{Qifa Yan}
\IEEEauthorblockA{ LTCI, T\'el\'ecom ParisTech\\ %\\Universite Paris-Saclay,\\
75013 Paris, France\\
Email: qifa.yan@telecom-paristech.fr}
\and
\IEEEauthorblockN{Sheng Yang\\   }
\IEEEauthorblockA{L2S, CentraleSup\'elec\\
%3 rue Joliot-Curie,\\
91190 Gif-sur-Yvette, France\\
Email: sheng.yang@centralesupelec.fr}
\and
\IEEEauthorblockN{Mich\`ele Wigger}
\IEEEauthorblockA{LTCI, T\'el\'ecom ParisTech \\
%Universite Paris-Saclay,\\
75013 Paris, France\\
Email:  michele.wigger@telecom-paristech.fr}
}

\maketitle

\IEEEpeerreviewmaketitle
%\begin{spacing}{2.0}
\begin{abstract} This paper investigates distributed computing systems where computations are split into ``Map" and ``Reduce" functions.
A new coded scheme, called  \emph{distributed computing and coded communication (D3C)}, is proposed, and its communication load is analyzed as a function of the available storage space and the number of intermediate values (IVA) to be computed. D3C achieves the smallest possible communication load for a given storage space, while a smaller number of IVAs need to be computed compared to  Li \emph{et al.'}s coded distributed computing (CDC) scheme. More generally, our scheme can flexibly trade between storage space and the number of IVAs to be computed. Communication load is then analyzed for any given tradeoff.
\end{abstract}

\section{Introduction}
Distributed computing has emerged as one of the most important approaches to big data analysis, where data-parallel computations are executed accross clusters consisting of  many  machines.  Platforms like MapReduce \cite{MapReduce} or Dryad \cite{Dryad}, for example, can perform computation tasks that involve data sets with tens of terabytes and more. In these systems, computations are distributed in the following way. Each cluster (referred to as \emph{node} in the sequel) $k\in\{1,\ldots, K\}$ is assigned the computation of a specific output function \begin{IEEEeqnarray}{c}\label{eq:decomposition}
	\phi_k(w_1,\cdots,w_N)=h_k(g_{k,1}(w_1),\cdots,g_{k,N}(w_N)),
\end{IEEEeqnarray}
which can depend on all data elements $w_1,\ldots, w_N$ but decomposes into smaller \emph{map functions} $g_{k,1},\ldots, g_{k,N}$ each depending only on a single data element. The computation of the output functions $\phi_1,\ldots, \phi_K$ is  carried out in three phases. During the first \emph{map phase},  each
 node $k$ locally stores a subset of the input data $\mathcal{M}_k\subseteq \{w_1,\ldots, w_N\}$ and calculates all
 intermediate values (IVAs) that depend on the stored data:
 \begin{equation}
 \{ g_{\ell,i}(w_i) \colon \; \ell \in \{1,\ldots, K\}, \; i \in \mathcal{M}_k\}.\notag
 \end{equation}
 In the subsequent \emph{shuffling phase},  the  nodes exchange the
 IVAs computed during the map phase. The goal is that  at the end of this second phase, each node $k$ is aware of all  IVAs $g_{k,1}(w_1),\ldots, g_{k,N}(w_N)$ required to calculate its output function $\phi_k$.  During the final \emph{reduce phase}, each node $k$ then
 calculates  the output function $\phi_{k}$  as in \eqref{eq:decomposition} based on the previously collected  IVAs  and the \emph{reduce function} $\phi_k$.

% In another line, coding has been shown to be powerful in reducing the communication load in caching networks. The coded caching technique, which  was originally proposed  by Maddah-Ali and Niesen to explore the \emph{multicast} opportunities created by the cached contents \cite{Maddah2014Fundamental,Maddah2015Decentralized},   has been extended to various settings such as device-to-device wireless networks \cite{Ji2016D2D}, online file update \cite{Pedarsani2016online,Yan2017online}, multiple antennas \cite{Ngo2018Scalable}, placement delivery arrays \cite{Yan2017PDA,Yan2018bipartite}, Gaussian broadcast channels \cite{Bidokhti2017Gaussian} \emph{et al.}.

 Li \emph{et al.} \cite{Li2018Tradeoff} showed that the communication cost  of such a distributed computation system can significantly be reduced if the nodes exploit  multicast coding opportunities during the  shuffling phase. More specifically, they designed a  coding scheme for the map and  shuffling phases, termed \emph{coded distributed computation (CDC)}, and proved that it has smallest communication load  among all distributed computation schemes with same storage  requirements.  Subsequently, various works extended the results in \cite{Li2018Tradeoff}. For example, the works in \cite{Lee2018codes,Li2016unified} account for straggling nodes; \cite{Qian2017How} studies optimal allocation of computation resources; and  \cite{Li2017Framework}  extends the works to wireless networks.

In this work, we propose to refine the performance measure and the coding scheme by Li \emph{et al.} \cite{Li2018Tradeoff}. Specifically, we account also for the number of intermediate values (IVAs) that each of the nodes has to calculate. In this sense, we extend the \emph{storage-communication tradeoff} studied in \cite{Li2018Tradeoff} to a \emph{storage-computation-communication tradeoff} where we wish to characterize the smallest communication load required in the shuffling phase for  a given storage space and a given  number of  IVAs totally calculated during the map phase. Notice that Li \emph{et al.} \cite{Li2018Tradeoff} termed their tradeoff \emph{computation-communication tradeoff} because they assume that each node calculates all the IVAs that can be obtained from the data stored at that node, irrespective of whether these IVAs are used in the sequel or not. In this sense, the total number of calculated IVAs is really a measure of the total storage space available at the nodes, this is why we refer to  it as storage-communication tradeoff.  The main result of this present paper is a new coding scheme, that we term \emph{distributed computing and coded communication (D3C)}. Our scheme in particular achieves the storage-computation tradeoff in \cite{Li2018Tradeoff} but with a reduced number of calculated IVAs  compared to    \cite{Li2018Tradeoff}.  %Moreover, the scheme  shows that one can have  the same communication load  with smaller storage space but each node calculating more IVAs or with larger storage space but each node calculating fewer IVAs.

%n particular, the optimal computation load for a given total cache size is in accordance with that in  CDC scheme. However, the computation load required to achieve this optimal communication load in the new scheme is significantly lower than that in CDC scheme. We emphasis that, our result is not conflict with the coverse established in \cite{Li2018Tradeoff}, because we adopted different computation load measure.

%The remainder of this paper is organized as follows. Section \ref{sec:model} introduces the system model  and Section \ref{sec:result} presents the main result of the paper. The  scheme achieving this main result is described and analyzed in Section \ref{sec:scheme}. Finally, the paper is concluded  in Section \ref{sec:conclusion}.

\textbf{Notations:} Let $\mathbb{N}^+$ be the set of positive integers. For $m,n\in\mathbb{N}^+$, denote the $n$ dimensional vector space over the finite field with cardinality $2^m$ by $\mathbb{F}_{2^m}^n$, and the integer set $\{1,2,\cdots,n\}$ by $[n]$. Scalars are denoted by upper or lower case letters. Sets or subsets are denoted by calligraphic font, and a collection (set of set) will be denoted by bold font. The cardinality of a set $\mathcal{S}$ is denoted by $|\mathcal{S}|$.  The bitwise Exclusive OR (XOR) operation is denoted by $\oplus$.

\section{System Model}\label{sec:model}

We consider a system consisting of $K$ distributed computing nodes and
%.  Each node intends to compute a distinct function from
$N$ input files each of $F$ bits, where $K,N,F\in\mathbb{N}^+$.
Specifically, given any $N$ files
\begin{IEEEeqnarray}{c}
\mathcal{W}:=\{w_1,\cdots,w_N\},\quad w_n\in \mathbb{F}_{2^F},\forall~n\in[N],\notag
 \end{IEEEeqnarray}
 the goal of Node $k$ ($k\in[K]$) is to compute an output function $\phi_k:\mathbb{F}_{2^{F}}^N\rightarrow \mathbb{F}_{2^B}$, which maps all the files to a bit stream $u_k=\phi_k(w_1,\cdots,w_N)\in\mathbb{F}_{2^B}$ of length $B$ for some $B\in\mathbb{N}^+$.

Following the conventions of \cite{Li2018Tradeoff,Li2017Framework}, we assume that the computation of the output functions $\phi_k$ can be decomposed as:
\begin{IEEEeqnarray}{c}
\phi_k(w_1,\cdots,w_N)=h_k(g_{k,1}(w_1),\cdots,g_{k,N}(w_N)),\notag
\end{IEEEeqnarray}
where we define the functions $g$ and $h$ as follows.
\begin{itemize}
  \item The ``map" function
\begin{IEEEeqnarray}{c}
g_{k,n}: \mathbb{F}_{2^F}\rightarrow \mathbb{F}_{2^T},~k\in[K],~ n\in[N],\notag
\end{IEEEeqnarray}
maps the file $w_n$ into a binary intermediate value (IVA) of $T$ bits, i.e., $v_{k,n}\triangleq g_{k,n}(w_n)\in \mathbb{F}_{2^T}$ for some $T\in\mathbb{N}^+$.
  \item The ``reduce" function
\begin{IEEEeqnarray}{c}
  h_k: \mathbb{F}_{2^T}^N\rightarrow \mathbb{F}_{2^B}, ~k\in[K],\notag
\end{IEEEeqnarray}
  maps the intermediate values
  \begin{IEEEeqnarray}{c}
  \mathcal{V}_k\triangleq\{v_{k,n}:n\in[N]\}\notag
  \end{IEEEeqnarray}
   of the output function $\phi_k$  into the output stream $u_k=h_k(v_{k,1},\cdots,v_{k,N})$.
\end{itemize}

Note that, such a decomposition always exists. For example,  one trivial decomposition is setting the map functions  be identity functions and  the reduce functions be the output functions, i.e., $g_{k,n}(w_n)=w_n$, and $h_k=\phi_k$, $\forall~n\in[N], k\in[K]$.

  \subsubsection{\textbf{Map Phase}} A central controller assigns  a subset  of files $\mathcal{M}_k\subset \mathcal{W}$ to Node $k$, for all $k\in[K]$.   Having access to the files in $\mathcal{M}_k$,  Node $k$ computes a subset of IVAs $\mathcal{C}_{k,n}=\{v_{q,n}:q\in \Lambda_{k,n}\}$, where $\Lambda_{k,n}\subset[K]$ for each file $w_n\in\mathcal{M}_k$. Denote the set of IVAs computed at Node $k$ by $\mathcal{C}_k$, i.e.,
\begin{IEEEeqnarray}{c}
    \mathcal{C}_k\triangleq \cup_{n:w_n\in\mathcal{M}_k}\mathcal{C}_{k,n}.\label{eqn:Ck}
\end{IEEEeqnarray}

To measure the storage and computation cost of the system, we introduce
the following two definitions.

\begin{definition}[Storage Space]
  We define the \emph{storage space}, denoted by $r$, as the total number of files stored across the $K$ nodes, normalized by the total number of files $N$, i.e.,
  \begin{IEEEeqnarray}{c}
    r\triangleq\frac{\sum_{k=1}^K|\mathcal{M}_k|}{N}.\label{eqn:r}
  \end{IEEEeqnarray}
\end{definition}

\begin{definition}[Computation Load] We define the computation load,
  denoted by $c$, as the total number of map functions computed across the  $K$ nodes,  normalized by the total number of map functions $NK$, i.e.,
  \begin{IEEEeqnarray}{c}
    c=\frac{\sum_{k=1}^K|\mathcal{C}_k|}{NK}.\label{eqn:c}
  \end{IEEEeqnarray}
\end{definition}

 \subsubsection{\textbf{Shuffle Phase}}  To compute the output function
 $\phi_k$, Node $k$ needs to collect the IVAs required by $\phi_k$ but
 not computed locally in the map phase, i.e., $\mathcal{V}_k\backslash \mathcal{C}_k$. After the map phase, the  $K$ nodes proceed to exchange the required IVAs. To be precise, each node $k$ creates a signal $X_k\in\mathbb{F}_{2^{l_k}}$ for some $l_k\in\mathbb{N}$, as a function of the IVAs computed in the map phase, i.e., for some encoding function
\begin{IEEEeqnarray}{c}
       \varphi_k: \mathbb{F}_{2^T}^{|\mathcal{C}_k|}\rightarrow
       \mathbb{F}_{2^{l_k}}.\notag
\end{IEEEeqnarray}
       Node $k$ multicasts the signal
\begin{IEEEeqnarray}{c}
       X_k=\varphi_k\left(\mathcal{C}_k\right),\notag
\end{IEEEeqnarray}
        to all the other nodes. Then we assume that each node $k$
        receives the signals $\{X_i\}_{i\in[K]\backslash\{k\}}$ without
        error.
        \begin{definition}[Communication Load] We define the communication load $L$, as the total length of the bits transmitted by the $K$ nodes during the shuffle phase normalized by the total length of all intermediate values $NKT$, i.e.,
\begin{IEEEeqnarray}{c}
        L\triangleq\frac{\sum_{k=1}^K l_k}{NKT}.\notag
\end{IEEEeqnarray}
        \end{definition}

\subsubsection{\textbf{Reduce Phase}}   With the signals $\{X_i\}_{i\in[K]}$, exchanged in the shuffle phase and the IVAs $\mathcal{C}_k$ computed locally, Node $k$ constructs all the IVAs $\mathcal{V}_k$ using some decoding function
  \begin{IEEEeqnarray}{c}
       \psi_k: \mathbb{F}_{2^{l_1}}\times\mathbb{F}_{2^{l_2}}\times\cdots\mathbb{F}_{2^{l_K}}\times \mathbb{F}_{2^T}^{|\mathcal{C}_k|}\rightarrow \mathbb{F}_{2^{T}}^{N},\notag
\end{IEEEeqnarray}
      So  Node $k$  computes
  \begin{IEEEeqnarray}{c}
       (v_{k,1},\cdots,v_{k,N})=\psi_k\left(X_1,\cdots,X_K,\mathcal{C}_k\right),\notag
\end{IEEEeqnarray}
       followed by the reduce function
 \begin{IEEEeqnarray}{c}
       u_k=h_k(v_{k,1},\cdots,v_{k,N}).\label{eqn:reducefunction}
\end{IEEEeqnarray}

\begin{definition}\label{def:SCC}
%For a  distributed computing system with $K$ nodes, a computation-cache-communication (C-C) triple $(c,r,L)$ is feasible if for any $\epsilon>0$, for sufficient large $N$, there exist $K$ file subsets $\{\mathcal{M}_k\}_{k=1}^K$, and its corresponding IVA sets $\{\mathcal{C}_k\}_{k=1}^K$, $K$ encoding functions $\{\varphi_k\}_{k=1}^K$ and  decoding functions $\{\psi_{k}\}_{k=1}^K$ such that, with a computation not exceeding $c$,  normalized cache size not exceeding $r$, and the  communication load  less than $L+\epsilon$,  each node $k$ ($k\in[K]$) can decode all the IVAs of its output function $\phi_k$ correctly, so that  it can proceed to compute the output function in the reduce phase.
A distributed computing system  achieves a
storage-computation-communication (SCC) triple $(r,c,L)$, if for any $\epsilon>0$, when
$N$ is sufficiently large, there exists a map-shuffle-reduce procedure such that  the  storage space, computation and communication loads do
not exceed $r+\epsilon$, $c+\epsilon$, and $L+\epsilon$, respectively. % each node $k$ ($k\in[K]$) can decode all the IVAs of its output function $\phi_k$ correctly, so that  it can proceed to compute the output function in the reduce phase.
In particular,  define the optimal communication load  as
\begin{IEEEeqnarray}{c}
L^*(r,c)\triangleq\inf\big\{L:(r,c,L)~\mbox{is achievable}\big\}.
\label{eq:Lcr}
\end{IEEEeqnarray}
\end{definition}
Note that it is without loss of generality to consider the case
\begin{align}
1\leq c\leq r<K.\notag
\end{align}%
Indeed, as $|\mathcal{C}_k|\leq |\mathcal{M}_k|K$ by \eqref{eqn:Ck}, thus $c\leq r$ from \eqref{eqn:r} and \eqref{eqn:c}.
Further, since each IVA needs to be computed at least once somewhere in
the system, we have $c\ge 1$. Finally, if
 $r\geq K$, each node  stores all files and can
 locally compute all the IVAs required for its output function.

%For each $c,r\in\mathbb{R}^+$, such that, $1\leq c\leq r<K$, define the optimal communication load by
%\begin{IEEEeqnarray}{c}
%L^*(c,r)\triangleq\inf\big\{L:(c,r,L)~\mbox{is a feasible C-C triple}\big\}.
%\end{IEEEeqnarray}

%We say that a caching-computation-communication tuple $(r,s,L)\in\mathbb{R}^3$ is feasible if for a system with normalized caching size $r$ and sufficient large $N$, there exist a set of feasible intermediate value sets $\{\mathcal{V}_k\}_{k=1}^K$, a set of encoding functions $\{\varphi_k\}_{k=1}^K$ and a set of decoding functions $\{\psi_{k}\}_{k=1}^K$ that achieve a computation load $\tilde{s}$, and communication load $\tilde{L}$, such that $\tilde{s}\leq s$, and $\tilde{L}\leq L$. For a given caching-computation tuple $(r,s)\in\mathbb{R}^2$, the optimal computation  load is defined as
%\begin{equation}
%L^*(r,s)\triangleq\inf\{L:(r,s,L)~
%\mbox{is feasible}\}.
%\end{equation}

 %In the rest of the paper, we are interested in identifying the optimal communication load $L^*(c,r)$  for all $1\leq c\leq r<K$  for any system with  $K\in\mathbb{N}^+$ nodes.

\begin{remark}\label{remark1} If each Node $k$ computes the IVAs required by all
  output functions from all the files
 that it stores, i.e., $\mathcal{C}_k=\{v_{q,n}:q\in[K], w_n\in\mathcal{M}_k\}$, then $|\mathcal{C}_k|=|\mathcal{M}_k|\cdot K$,  and
\begin{IEEEeqnarray}{rCl}
c&=&\frac{\sum_{k=1}^K|\mathcal{C}_k|}{NK}
=\frac{\sum_{k=1}^K|\mathcal{M}_k|\cdot K}{NK}=r.\notag
\end{IEEEeqnarray}

This is exactly the case investigated in \cite{Li2018Tradeoff}, in which
the authors established the fundamental storage-communication
tradeoff. The \emph{optimal} tradeoff curve is given by  the lower convex envelope of
$
\left\{\left(r,L^*(r)\right):r=1,\cdots,K\right\}
$,
where
\begin{IEEEeqnarray}{c}
  L^*(r) \triangleq \frac{1}{r}\left(1-\frac{r}{K}\right). \label{eq:L*r}
\end{IEEEeqnarray}
In other words, the optimal computation load  in
\eqref{eq:Lcr} is known for the case $c=r$ where $L^*(r,r) =
L^*(r)$. Intuitively, however,
computing the IVAs for all other nodes may be highly redundant across
the whole system. This has motivated our investigation in
the more general case where the computation load $c$ can be strictly smaller
than the storage space $r$.

\end{remark}

\section{Main Result}\label{sec:result}

\begin{theorem}\label{thm:main}In a distributed computing system with
  $K$ nodes, for any storage space $r\in[1,K)$,  and
\begin{IEEEeqnarray}{c}
c\in\bigg\{\frac{r}{K}+\left(1-\frac{r}{K}\right)g~:~g=1,2,\cdots,\lfloor r\rfloor\bigg\},\label{eqn:c_range}
\end{IEEEeqnarray}
  the following communication load $L(r,c)$ is achievable:
 \begin{IEEEeqnarray}{rCl}
 L(r,c)=\frac{1}{c-r/K}\cdot\left(1-\frac{r}{K}\right)^2.\label{eqn:optimalL}
 \end{IEEEeqnarray}
 Define
 \begin{IEEEeqnarray}{c}
c^*(r)\triangleq\frac{r}{K}+\left(1-\frac{r}{K}\right)\cdot g_r,\notag
\end{IEEEeqnarray}
  where
  \begin{IEEEeqnarray}{c}
  g_r\triangleq \lfloor r\rfloor +\frac{(r-\lfloor r\rfloor)(K-\lceil r\rceil)}{K-r}.\label{eqn:gr}
  \end{IEEEeqnarray}
  For $c^*(r) \leq c\leq r$, $L^*(r)$ in \eqref{eq:L*r} is achievable.

  For general $c\in\left[1,c^*(r)\right]$, the lower convex envelope of
  \eqref{eqn:c_range}--\eqref{eqn:optimalL} and the point
  $\big(c^*(r),L^*(r)\big)$ is achievable, where $L^*(r)$ is given by
  \eqref{eq:L*r}.

\end{theorem}

 \begin{figure}%[htp!]
  \centering
  \includegraphics[width=0.4\textwidth,height=0.24\textwidth]{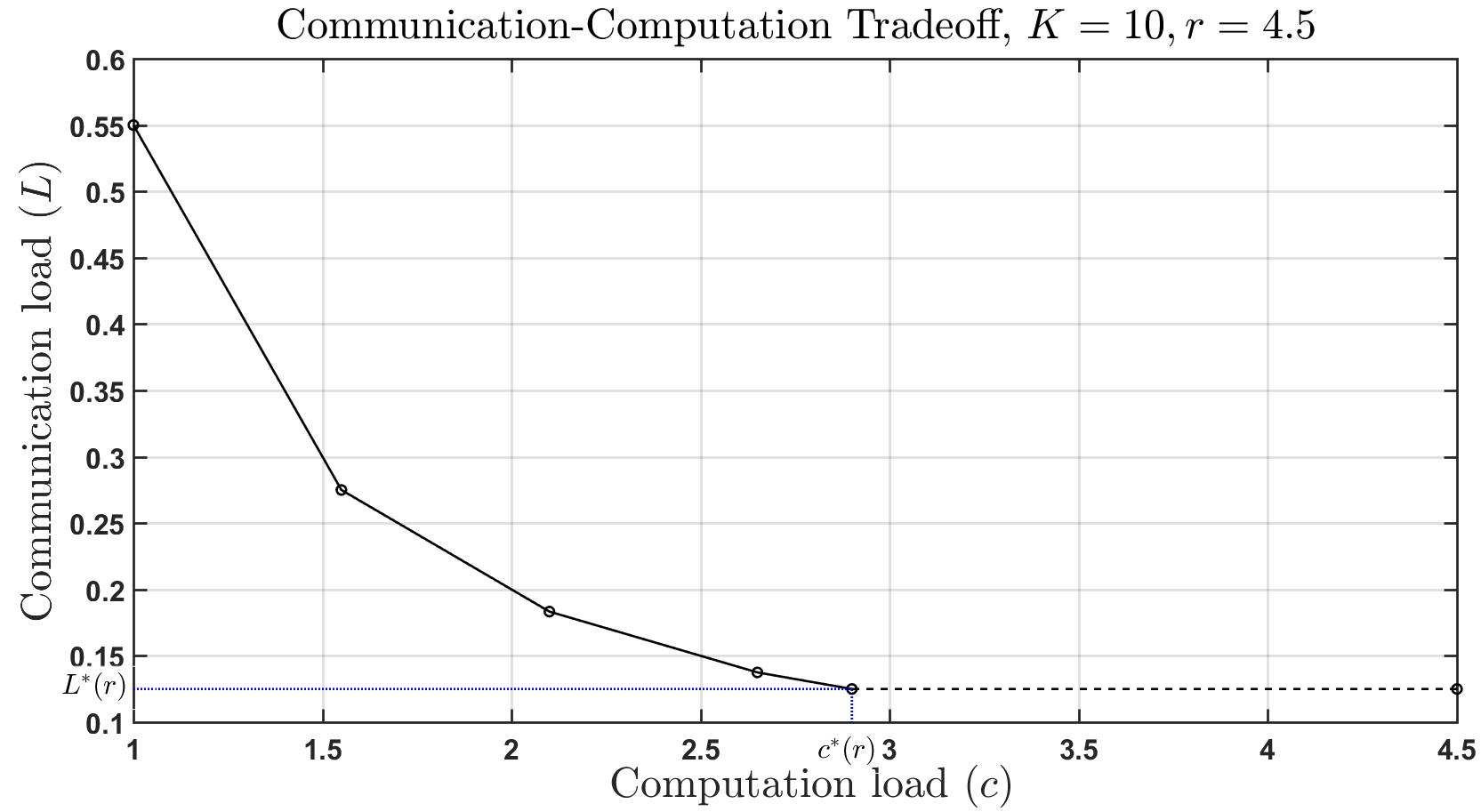}\\
  \caption{The communication-computation curve $L(r,c)$ for $K=10,r=4.5$. The corner points correspond to different values of $g$ as given in \eqref{eqn:c_range} and \eqref{eqn:gr}.}\label{Fig:Tradeoff}
\end{figure}

Notice that for any  fixed $r\in[1,K)$, the achievable communication
load $L(r,c)$ is a piecewise linear function over $c\in[0,r]$. Fig.
\ref{Fig:Tradeoff} illustrates the case $K=10,~ r=4.5$. When $c$
increases from $1$ to $c^*(r)$, the communication load decreases from
$1-\frac{r}{K}$ to $L^*(r)$. The communication load becomes constant
when $c\ge c^*(r)$. When $c\in[c^*(r),r]$,  from
Theorem~\ref{thm:main}, we have $L^*(r,c) \le L^*(r) = L^*(r,r)$, and
thus
\begin{IEEEeqnarray}{c}
  L^*(r,r) \ge L^*(r,c) \ge L^*(r,r),\notag
\end{IEEEeqnarray}
where the last inequality holds because $L^*(r,c)$ is (non-strictly) decreasing
 respect to $c$. Hence, we establish the following
optimality.
 \begin{corollary}
 In a distributed computing system with $K$ nodes %, %$K\in\mathbb{N}^+$,
 and storage space $r\in[1,K)$, for $c\in[c^*(r),r]$,
\begin{IEEEeqnarray}{c}
L^*(r,c)=L^*(r).\notag
\end{IEEEeqnarray}
\end{corollary}
%\begin{proof}
%i.e., $L^*(r,r)$ is characterize by the convex envelop of the points in \eqref{eqn:envelop}. That is, let  $\eta=r-\lfloor r\rfloor$, satisfies
%\begin{IEEEeqnarray}{c}
%r=(1-\eta)\cdot\lfloor r\rfloor +\eta\cdot\lceil r\rceil,
%\end{IEEEeqnarray}
%Then,
%\begin{IEEEeqnarray}{rCl}
%L^*(r)&\triangleq&L^*(r,r)\\
%&=& (1-\eta) \frac{1}{\lfloor r \rfloor}\big(1-\frac{\lfloor r\rfloor}{K}\big)+\eta\frac{1}{\lceil r\rceil}\big(1-\frac{\lceil r\rceil}{K}\big)\\
%&=&\frac{\lceil r\rceil+\lfloor r\rfloor -r}{\lceil r\rceil \lfloor r\rfloor}-\frac{1}{K}\label{eqn:Lsm}
%\end{IEEEeqnarray}
%\end{proof}

Thus, for  fixed $r\in[1,K)$, to achieve the
optimal communication load $L^*(r)$,  computation load $c^*(r)$
suffices. This computation load is significantly lower than when each node computes  IVAs for all output functions from all its stored files (Remark \ref{remark1}). The following example illustrates this saving.
 \begin{example}
Let $K=2,$ $r=2$, and $N=6$. Consider Fig. \ref{Fig:D3Cscheme}. The top-most line in each of the three boxes indicates the  files stored at the node. Below this line, the computed IVAs are depicted, where red circles indicate IVAs $\{v_{1,1}, \ldots, v_{1,6}\}$, green squares IVAs $\{v_{2,1}, \ldots, v_{2,6}\}$, and blue triangles  IVAs $\{v_{3,1}, \ldots, v_{3,6}\}$.  Nodes $1$, $2$ and $3$ need to collect the IVAs denoted by the circles, square and triangles respectively. The last line of each box indicates the  IVAs that the node needs to learn during the shuffling phase.  %the ap and Shuffling is exactly the same as in  CDC scheme (See \cite[Example $1$]{Li2018Tradeoff}).

 The dashed circles/squares/triangles stand for the IVAs that would be computed in addition in the CDC scheme, see \cite[Example $1$]{Li2018Tradeoff}. In other words, they represent the saving in computation load. In fact for this example, D3C and CDC have computation loads  $\frac{4}{3}$ and $2$, respectively.
\begin{figure}%[htp!]
  \centering
  \includegraphics[width=0.4\textwidth,height=0.24\textwidth]{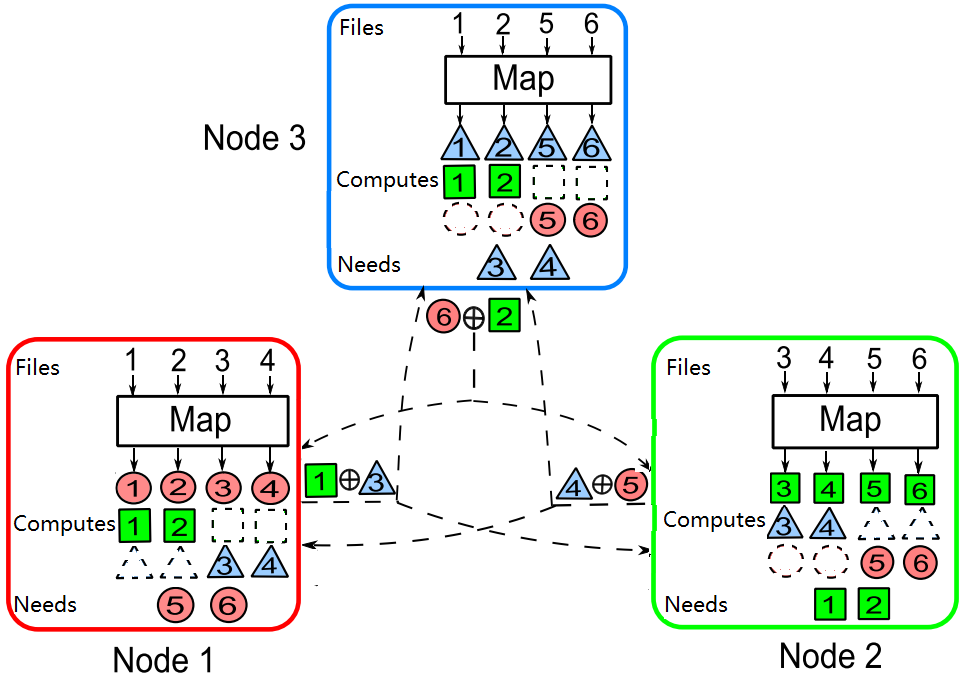}\\
  \caption{ D3C scheme achieving $L^*(r)$, for a system with $K=3, r=2$. }\label{Fig:D3Cscheme}
\end{figure}
\end{example}

% \begin{enumerate}
%   \item For $1\leq c\leq \rho_r$, the optimal $L^*(c)$ is completely characterized;
%   \item For $\rho_r< c\leq \tau_r$, an universal upper bound for the multiplicative gap of $L(c)$ and $L^*(c)$ can be obtained by observing that $\lceil r\rceil\geq 2$:
%\begin{IEEEeqnarray}{c}
%\frac{L(c)}{L^*(c)}\leq 1.125.
%\end{IEEEeqnarray}
%As indicated by \eqref{eqn:gap}, this gap improves with the increase of $r$. For example, if $r>2$, the gap does not exceed $1.042$.
%   \item When $c\geq \tau_r$, $L^*(c)$ can't be further reduced when $c$ increases.
% \end{enumerate}
%
%Denote the righthand side (RHS) of \eqref{eqn:Lminstar} by
%   \begin{IEEEeqnarray}{rCl}
%&&L_{\min}\\
%&\triangleq&\frac{\lceil r\rceil+\lfloor r\rfloor-r}{\lceil r\rceil \lfloor r\rfloor}-\frac{1}{K}\\
%&=&(1-\eta)\frac{1}{\lfloor r\rfloor}\big(1-\frac{\lfloor r\rfloor}{K}\big)+\eta\frac{1}{\lceil r\rceil}\big(1-\frac{\lceil r\rceil}{K}\big)
%\end{IEEEeqnarray}
%   where
%   $
%   \eta\triangleq r-\lfloor r\rfloor
%   $
%   satisfies
%   \begin{IEEEeqnarray}{c}
 %  r=(1-\eta)\cdot \lfloor r\rfloor+\eta\cdot \lceil r\rceil.
%   \end{IEEEeqnarray}

\begin{figure}%[htp!]
  \centering
  \includegraphics[width=0.4\textwidth,height=0.24\textwidth]{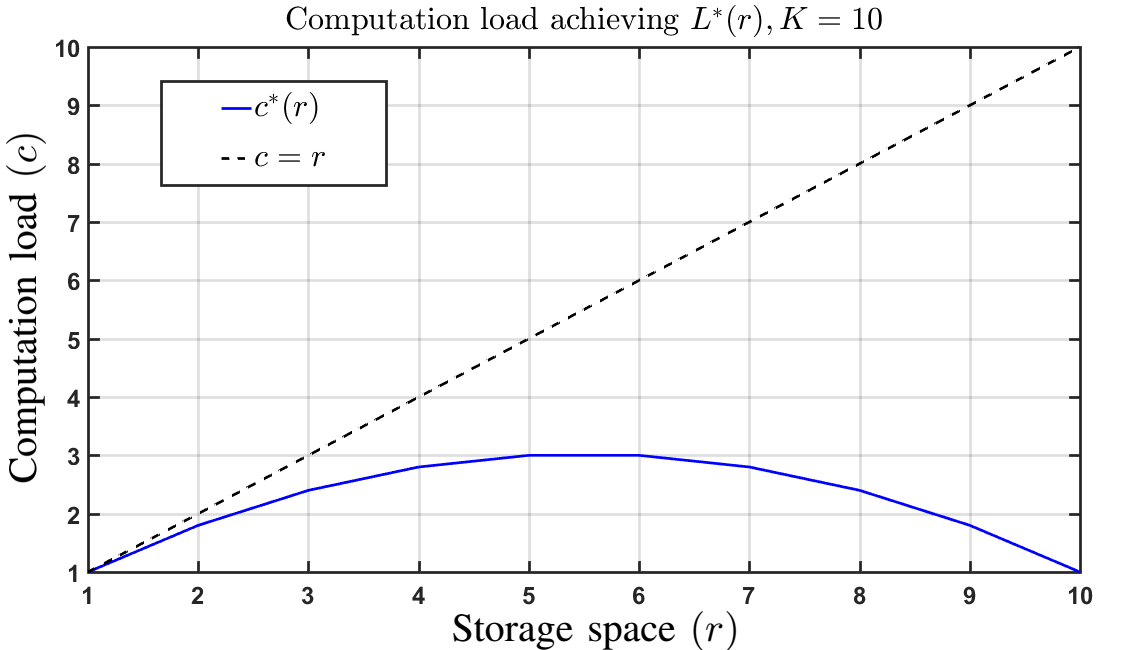}\\
  \caption{The computation load achieving $L^*(r)$.}\label{Fig:IVAs}
\end{figure}

To visualize savings in computation load,   we plot
$c^*(r)$ as a function of the  storage space $r$ as well as
the line $c=r$  in Fig. \ref{Fig:IVAs}.  It implies that  when the storage space exceeds a certain threshold, the
necessary computation load is decreasing in the storage space. This is because with
larger storage space, more IVAs can be locally computed  by the nodes requesting them, thus each node computes less IVAs  for coded communication, which consumes more computation.

  We will prove Theorem \ref{thm:main} by proposing a distributed
  computing scheme, termed \emph{ditributed computing and coded
  communication} (D3C) scheme. For $c=c^*(r)$, the
shuffle phase of D3C scheme specializes to the shuffle phase of the CDC
  scheme. But the map phase of  our D3C scheme only computes those IVAs
  that are necessary for communication  and decoding. This special case has been investigated in \cite{Ezzelldin2017ITW}, and similar conclusions are  obtained there.

%Theorem \ref{thm:main} characterizes an achievable tradeoff between the computation load and communication load: when $c$ increases from $1$ to $c_{\max}$, where
%\begin{IEEEeqnarray}{rCl}
%c_{\max}&\triangleq&r-\lfloor r\rfloor \cdot\frac{2r-1-\lfloor r\rfloor}{K},\label{eqn:smax}
%\end{IEEEeqnarray}
%the communication load decreases from $1-\frac{r}{K}$ to $L_{\min}$, where
%\begin{IEEEeqnarray}{rCl}
%L_{\min}&\triangleq&(1-\theta_{\max})\cdot\frac{1}{\lfloor r\rfloor}\cdot\big(1-\frac{\lfloor r\rfloor}{K}\big)\\
%&&+\theta_{\max}\cdot\frac{1}{\lceil r\rceil}\cdot\big(1-\frac{\lceil r\rceil}{K}\big)\label{eqn:Lmin}\\
%&=&\frac{\lceil r\rceil+\lfloor r\rfloor-r}{\lceil r\rceil \lfloor r\rfloor}-\frac{1}{K}.
%\end{IEEEeqnarray}

%$\frac{r}{K}+\big(1-\frac{r}{K}\big)\cdot\big(\lfloor r\rfloor +(\lceil r\rceil -r)\cdot\frac{K-\lceil r\rceil}{K-r}\big)$.

\section{Distributed Computing and Coded Communication Scheme}\label{sec:scheme}

In this section, we describe the proposed scheme that achieves the
result in Theorem~\ref{thm:main}.  Define
\begin{IEEEeqnarray}{c}
g=\frac{c-r/K}{1-r/K}.\notag
\end{IEEEeqnarray}
We will first
present the scheme for the case $r\in\{1,\cdots, K\}$, and
 $g\in\{1,\cdots,r\}$, which
will be referred to as the basic D3C scheme hereafter. Then we will extend the basic D3C scheme to obtain the achievability for other cases.

 For   the case $r\in\{1,\cdots,K\},  g\in\{1,\cdots,r\}$, the $N$ input files are partitioned into ${K\choose r}{r\choose g}$ batches, each containing
\begin{IEEEeqnarray}{c}
\eta_g:=\frac{N}{{K\choose r}{r\choose g}}\label{eqn:eta}
\end{IEEEeqnarray}
files. Associate each batch with an element $(\mathcal{S},\mathcal{T})$ of the set
\begin{IEEEeqnarray}{rCl}\label{eq:conditions}
\mathbf{\Omega}:= \big\{(\mathcal{S},\mathcal{T}) \colon \; \mathcal{T}\subset \mathcal{S}\subset[K], \; \; |\mathcal{S}|=r,\;\; |\mathcal{T}|=g\big\},\IEEEeqnarraynumspace\notag
\end{IEEEeqnarray}
and let $\mathcal{W}_{\mathcal{S},\mathcal{T}}$ denote the batch of $\eta_g$ files associated
to a pair $(\mathcal{S},\mathcal{T})$. Notice that
\begin{IEEEeqnarray}{rCl}
\mathcal{W}&:=&\{w_1,\cdots,w_N\} = \bigcup_{(\mathcal{S},\mathcal{T})\in\mathbf{\Omega}}\mathcal{W}_{\mathcal{S},\mathcal{T}}.\notag
\end{IEEEeqnarray}
 Let further $\mathcal{U}_{k,\mathcal{S},\mathcal{T}}$ be the set of IVAs
for output function $\phi_k$ that can be computed from the files in
$\mathcal{W}_{\mathcal{S},\mathcal{T}}$:
\begin{IEEEeqnarray}{c}
\mathcal{U}_{k,\mathcal{S},\mathcal{T}}\triangleq\{v_{k,n} \colon n\in [N] \textnormal{ so that } w_n\in \mathcal{W}_{\mathcal{S},\mathcal{T}}\}.\notag
\end{IEEEeqnarray}

We now describe the three distributed-computing  phases.

1) \emph{Map phase:} Node $k$ stores the files in $\mathcal{W}_{\mathcal{S},\mathcal{T}}$ if and only if $k \in \mathcal{S}$. Thus:
\begin{IEEEeqnarray}{c}
\mathcal{M}_k=\bigcup_{\substack{(\mathcal{S},\mathcal{T})\in \mathbf{\Omega} \colon\\  k\in \mathcal{S}} }\mathcal{W}_{\mathcal{S},\mathcal{T}}.\notag
\end{IEEEeqnarray}
Node $k$ further computes the IVAs in
\begin{IEEEeqnarray}{c}
\mathcal{C}_k=\mathcal{C}_k^1\cup\mathcal{C}_k^2,\label{eqn:Vk}
\end{IEEEeqnarray}
where
\begin{IEEEeqnarray}{rCl}
\mathcal{C}_k^1&:=&\bigcup_{\substack{(\mathcal{S},\mathcal{T})\in \mathbf{\Omega} \colon\\  k\in \mathcal{S}} }  \mathcal{U}_{k,\mathcal{S},\mathcal{T}},\label{eqn:Vk1}\\
\mathcal{C}_k^2&:= &\bigcup_{\substack{(\mathcal{S},\mathcal{T})\in \mathbf{\Omega} \colon\\  k\in \mathcal{T}} } \bigcup_{q \in [K]\backslash \mathcal{S}} \;\;
 \mathcal{U}_{q,\mathcal{S},\mathcal{T}}.\label{eqn:Vk2}
\end{IEEEeqnarray}
It's easy to verify that all the IVAs in $\mathcal{C}_k$ can be computed from  $\mathcal{M}_k$, i.e., the files stored at Node $k$.

2) \emph{Shuffling Phase:} For each element $(\mathcal{S},\mathcal{T})\in\mathbf{\Omega}$ and each index $j\in [K] \backslash \mathcal{S}$, we partition
 %In particular, the IVAs $\mathcal{U}_{j,\mathcal{S},\mathcal{T}}$ ($j\notin \mathcal{S}$) is computed by each user $k\in \mathcal{T}$. Accordingly, we partition each
 the set $\mathcal{U}_{j,\mathcal{S},\mathcal{T}}$ into $g$ equal smaller subsets
 \begin{equation}
\mathcal{U}_{j,\mathcal{S},\mathcal{T}}= \left\{ \mathcal{U}_{j,\mathcal{S},\mathcal{T}}^k\colon \; k \in \mathcal{T}\right\}.\notag
 \end{equation} %of equal size.
%\begin{IEEEeqnarray}{rCl}
%\mathcal{U}_{k,\mathcal{S},\mathcal{T}}&=&\bigcup_{j\in \mathcal{T}}\mathcal{U}_{k,\mathcal{S},\mathcal{T}}^j,\label{eqn:Vjst}\\
% \mathcal{U}_{k,\mathcal{S},\mathcal{T}}^{j_1}\cap \mathcal{U}_{k,\mathcal{S},\mathcal{T}}^{j_2}&=&\emptyset,\qquad \forall~j_1,j_2\in \mathcal{T},~\; j_1\neq j_2.
%\end{IEEEeqnarray}
Define now the set
\begin{IEEEeqnarray}{rCl}\label{eq:conditions}
	\mathbf{\Pi}:= \big\{(\mathcal{I},\mathcal{J}) \colon \; \mathcal{J}\subset \mathcal{I}\subset[K], \; \; |\mathcal{I}|=r+1,\;\; |\mathcal{J}|=g+1\big\}.\nonumber
\end{IEEEeqnarray}
For each  $(\mathcal{I},\mathcal{J})\in\mathbf{\Pi}$ and $k\in \mathcal{J}$, by \eqref{eqn:Vk2},   Node $k$ can compute
the signal
\begin{IEEEeqnarray}{c}
X_{\mathcal{I}, \mathcal{J}}^{k}:=\bigoplus_{i\in \mathcal{J}\backslash\{k\}}\mathcal{U}_{i,\mathcal{I}\backslash\{i\}, \mathcal{J}\backslash\{i\}}^k\notag
\end{IEEEeqnarray}
from the IVAs calculated during the map phase.  A given Node $k$ thus sends (shuffles) the multicast signal
 \begin{IEEEeqnarray}{rCl}
 	X_k = \big\{ X_{\mathcal{I}, \mathcal{J}}^{k} \colon \; ( \mathcal{I}, \mathcal{J}) \in \mathbf{\Pi} \; \textnormal{such that } k\in  \mathcal{J} \big\}.\notag
 \end{IEEEeqnarray}

3) \emph{Reduce Phase:} Notice that $\mathcal{C}_k^{2}$ only contains IVAs $v_{q,n}$ where $q \neq k$. Thus, by  \eqref{eqn:Vk}--\eqref{eqn:Vk1},  during the shuffling phase  each node $k$ needs to learn  all IVAs in
\begin{IEEEeqnarray}{c}
	\bigcup_{\substack{(\mathcal{S},\mathcal{T})\in \mathbf{\Omega} \colon\\  k\notin \mathcal{S}} }  \mathcal{U}_{k,\mathcal{S},\mathcal{T}}.\notag
\end{IEEEeqnarray}

Fix an arbitrary pair $(\mathcal{S},\mathcal{T})\in \mathbf{\Omega}$ such that $k \notin \mathcal{S}$ and an element $j\in \mathcal{T}$. From the received multicast message $X_{\mathcal{S}\cup \{k\}, \mathcal{T}\cup \{k\}}^{j}$ during the shuffling phase and  its locally calculated IVAs during the map phase
\begin{IEEEeqnarray}{rCl}
\Big\{ \mathcal{U}_{i,\mathcal{S}\cup\{k\}\backslash\{i\},\mathcal{T}\cup\{k\}\backslash\{i\}}^j \colon \; \;  i \in \mathcal{T}\backslash \{j\}\Big\}, \IEEEeqnarraynumspace\notag
\end{IEEEeqnarray}
Node $k$ can recover the missing IVAs   $\mathcal{U}_{k,\mathcal{S},\mathcal{T}}^j$ through a simple XOR operation.
% Receiving the signals in the shuffle phase, for each $j\in [K],~(\mathcal{S},\mathcal{T})$ satisfying \eqref{eq:conditions} such that $j\notin \mathcal{S}$, Node $j$ receives  $\mathcal{U}_{k,\mathcal{S},\mathcal{T}}$.
%\begin{IEEEeqnarray}{c}
%\bigoplus_{i\in \mathcal{T}\cup\{k\}\backslash\{i\}} \mathcal{U}_{i,\mathcal{S}\cup\{k\}\backslash\{i\},\mathcal{T}\cup\{j\}\backslash\{i\}}^j
%\end{IEEEeqnarray}
%from each Node $k\in \mathcal{T}$. Thus, for each $k\in \mathcal{T}$,  with \eqref{eqn:Vk2} and \eqref{eqn:Vjst}, Node $j$ has computed all IVAs $\mathcal{U}_{i,\mathcal{S}\cup\{j\}\backslash\{i\},\mathcal{T}\cup\{j\}\backslash\{i\}}^k$ $(i\neq j,~ i\neq k)$.
 After collecting all missing IVAs, Node $k$ can proceed to compute the reduce function \eqref{eqn:reducefunction}.

4) \emph{Analysis:}
 We  analyze the performance of  the scheme.
\begin{enumerate}
  \item \emph{Storage Space:} The number of batches in $\mathcal{M}_k$ is
$
{K-1\choose r-1}{r\choose g}\notag
$,
each consisting of  $\eta_g $ files. Thus, the  storage space is
\begin{IEEEeqnarray}{rCl}
\frac{1}{N}\cdot K\cdot{K-1\choose r-1}{r\choose g}\cdot\eta_g %&=&{K-1\choose r-1}\cdot{r\choose g}\cdot\frac{N}{{K\choose r}{r\choose g}}\cdot F\\
&=&r.\notag
\end{IEEEeqnarray}

  \item\emph{Computation Load:} %For each $k\in[K]$, notice that
  Since $\mathcal{C}_k^1\cap \mathcal{C}_k^2=\emptyset$: %, and $|\mathcal{U}_{k,\mathcal{S},\mathcal{T}}|=|\mathcal{W}_{\mathcal{S},\mathcal{T}}|=\eta_g$. Thus,
\begin{IEEEeqnarray}{rCl}
|\mathcal{C}_k|&=&|\mathcal{C}_{k}^1|+|\mathcal{C}_k^2|.\label{eqn:Va}
\end{IEEEeqnarray}
By \eqref{eqn:eta}, \eqref{eqn:Vk1}, \eqref{eqn:Vk2}, we have
\begin{IEEEeqnarray}{rCl}
|\mathcal{C}_{k}^1|&=&{K-1\choose r-1} {r\choose g}\cdot\eta_g=\frac{rN}{K},\label{eqn:Vb}\\
|\mathcal{C}_{k}^2|&=&{K-1\choose r-1}{r-1\choose g-1}\cdot(K-r)\cdot\eta_g\notag\\
&=&\left(1-\frac{r}{K}\right)\cdot g\cdot N.\label{eqn:Vc}
\end{IEEEeqnarray}
Thus, by \eqref{eqn:Va}, \eqref{eqn:Vb} and \eqref{eqn:Vc}, the computation load is
%\begin{IEEEeqnarray}{c}
%|\mathcal{V}_k|=\frac{rN}{K}+\left(1-\frac{r}{K}\right)\cdot g\cdot N\notag
%\end{IEEEeqnarray}
\begin{IEEEeqnarray}{rCCCl}
c&=&\frac{\sum_{k=1}^K|\mathcal{C}_k|}{NK}&=&\frac{r}{K}+\left(1-\frac{r}{K}\right)g.\label{eqn:computation}
\end{IEEEeqnarray}
  \item \emph{Communication Load:} The number of signals that each Node $k$ transmits is
${K-1\choose r}\cdot{r\choose g}$,
each of size $\frac{\eta_g\cdot T}{g}$ bits. Thus, the communication load is
\begin{IEEEeqnarray}{rCl}
L&=&\frac{\sum_{k=1}^K|X_k|}{NKT}\notag\\
&=&\frac{1}{NKT} \cdot K{K-1\choose r} {r\choose g}\frac{NT}{{K\choose r}{r\choose g}}\cdot\frac{1}{g}\notag\\
&=&\frac{1}{g}\cdot\left(1-\frac{r}{K}\right)\notag\\
&\overset{(a)}{=}&\frac{1}{c-r/K}\cdot\left(1-\frac{r}{K}\right)^2,\notag
\end{IEEEeqnarray}
where $(a)$ follows from \eqref{eqn:computation}.
\end{enumerate}

This concludes the achievability proof for $r\in\{1,\cdots,K\},$ $g\in\{1,\cdots,r\}$.
We complete the proof by proposing several extensions of the basic D3C scheme:
\begin{enumerate}
  \item[E$1$.] For any $r\in[1,K)$, and $g\in\{1,\cdots,\lfloor r\rfloor\}$, there exist an unique $\alpha\in[0,1)$ such that
\begin{IEEEeqnarray}{c}
r=(1-\alpha)\lfloor r\rfloor+\alpha \lceil r\rceil.\notag
\end{IEEEeqnarray}
We partition the $N$ files into two groups with cardinalities\footnote{While  $\alpha N$ and $(1-\alpha)N$ being integers requires that, $\alpha$ has to be a  rational number, irrational $\alpha$ can be approached arbitrarily close by a series of rational numbers, since the rational numbers are dense in the real interval $[0,1]$. As Definition \ref{def:SCC} allows a $\epsilon$-discrepancy, this does not affect the proof.} $(1-\alpha)N$ and $\alpha N$ respectively. For the first group, implement the basic D3C scheme with a  storage space $\lfloor r\rfloor$, and computing load $c_1\triangleq\frac{\lfloor r\rfloor}{K}+\big(1-\frac{\lfloor r\rfloor}{K}\big)g$. For the second group, implement the basic D3C scheme with a  storage space $\lceil r\rceil$, and computing load $c_2\triangleq\frac{\lceil r\rceil}{K}+\big(1-\frac{\lceil r\rceil}{K}\big)g$. %The the computation load $c_1$ and communication load $L_1$ are given by:
%\begin{IEEEeqnarray}{rCl}
%&&c_1\\
%&=&\frac{1}{NK}\cdot\left[\left(\frac{\lfloor r\rfloor}{K}+\left(1-\frac{\lfloor r\rfloor}{K}\right)\lfloor r\rfloor\right)\alpha NK\right.\notag\\
%&&\left.+\left(\frac{\lfloor r\rfloor}{K}+\left(1-\frac{\lfloor r\rfloor}{K}\right)\lfloor r\rfloor\right)(1-\alpha)NK\right]\\
%&=&\frac{r}{K}+\left(1-\frac{r}{K}\right)g.\\
%&&L_1\\
%&=&\frac{1}{NKT}\cdot\left[\frac{1}{g}\left(1-\frac{\lfloor r\rfloor}{K}\right)\cdot\alpha NKT\right.\notag\\
%&&\left.+\frac{1}{g}\left(1-\frac{\lceil r\rceil}{K}\right)\cdot(1-\alpha) NKT\right]\\
%&=&\frac{1}{g}\left(1-\frac{r}{K}\right).
%\end{IEEEeqnarray}
%The C-C triple $(\tilde{c},\tilde{r}, \tilde{L})$ achieved by this scheme is:
%\begin{IEEEeqnarray}{rCl}
%\tilde{r}&=&\frac{1}{N}\big(\lfloor r\rfloor\cdot(1-\alpha)N+\lceil r\rceil\cdot\alpha N\big)=r,\label{eqn:rtilde}\\
%\tilde{c}&=&\frac{1}{NK}\big(c_1\cdot (1-\alpha) NK+c_2\cdot\alpha NK\big)=c,\label{eqn:ctilde}\\
%\tilde{L}&\overset{(a)}{=}&\frac{1}{NKT}\bigg(\frac{1}{g}\big(1-\frac{\lfloor r\rfloor}{K}\big)\cdot (1-\alpha)NKT
%+\frac{1}{g}\big(1-\frac{\lceil r\rceil}{K}\big)\cdot \alpha NKT\bigg)\\
%&=&\frac{1}{g}\big(1-\frac{r}{K}\big)\\
%&\overset{(b)}{=}&\frac{1}{c-r/K}\cdot\big(1-\frac{r}{K}\big)^2,
%\end{IEEEeqnarray}
%where $(a)$ and $(b)$ follows from \eqref{eqn:g_perm} and \eqref{eqn:computation} respectively.
This results  the achievability of \eqref{eqn:optimalL},  when $r\in[1,K)$, $g\in\left\{1,\cdots,\lfloor r\rfloor\right\}$.
\item[E$2$.] For any $r\in[1,K)$ and for  $g\in\big[1,\lfloor r\rfloor\big)$,  there exists an unique $\beta\in[0,1)$ such that
    \begin{IEEEeqnarray}{rCl}
    g&=&(1-\beta)\lfloor g\rfloor+\beta\lceil g\rceil.\notag
    \end{IEEEeqnarray}
We partition the $N$ files into two groups with cardinalities $(1-\beta)N$ and $\beta N$ respectively. For the first group, implement the  scheme in E$1$ with a  storage space $r$, and computing load $c_1'\triangleq\frac{ r}{K}+\big(1-\frac{ r}{K}\big)\lfloor g\rfloor$. For the second group, implement the scheme in E$1$ with a  storage space $ r$, and computation load $c_2'\triangleq\frac{ r}{K}+\big(1-\frac{ r}{K}\big)\lceil g\rceil$. %Similar to \eqref{eqn:rtilde} and \eqref{eqn:ctilde}, the normalized cache size and computation load is $r$ and $c$ respectively. The communication load $\bar L$ is
%\begin{IEEEeqnarray}{rCl}
%\bar{L}&=&(1-\beta)\cdot\frac{1}{c_1-r/K}\big(1-\frac{r}{K}\big)^2
%+\beta\cdot\frac{1}{c_2-r/K}\big(1-\frac{r}{K}\big)^2.
%\end{IEEEeqnarray}
Notice that, $\beta$ satisfies $c=(1-\beta) c_1'+\beta c_2'$. Thus, this results  the achievability of the lower convex envelop in \eqref{eqn:optimalL} of Theorem \ref{thm:main}.
\item[E$3$.] When $r\notin\mathbb{N}$, $g\in\left[\lfloor r\rfloor, g_r\right]$, %. When $c\in(\rho_r,\tau_r]$\footnote{Note that, if $(\rho_r,\tau_r]\neq\emptyset$, i.e., $\tau_r>\rho_r$, then it indicates that $r$ is not an integer.},
    then there exists an unique $\theta\in(0,r-\lfloor r\rfloor]$ such that
    \begin{IEEEeqnarray}{c}
    c=\frac{r}{K}+\left(1-\frac{r}{K}\right)\cdot\left(\lfloor r\rfloor+\theta\cdot\frac{K-\lceil r\rceil}{K-r}\right).\notag
    \end{IEEEeqnarray}

    Define
    $
    r'(\theta)\triangleq r-\frac{\theta}{1-\theta}\cdot(\lceil r\rceil-r),\notag
   $
    or equivalently,
    \begin{IEEEeqnarray}{c}
    r=(1-\theta)\cdot r'(\theta)+\theta\cdot \lceil r\rceil.\notag
    \end{IEEEeqnarray}

 %When $r\in[K]$, $\theta=0$, so the achievability follows from the first part. Now we assume $r\in[1,K)\backslash[K]$, then $\lceil r\rceil -\lfloor r\rfloor=1$.
 Notice that $r'(\theta)$ decreases with $\theta$, so
    \begin{IEEEeqnarray}{rCl}
    r'(\theta)&\geq &\left.r'(\theta)\right|_{\theta=r-\lfloor r\rfloor}=\lfloor r\rfloor.\notag
    \end{IEEEeqnarray}
     We partition the $N$ files into two groups with cardinalities $(1-\theta)N$ and $\theta N$ respectively. For the first group, implement the scheme in E$1$ with a  storage space $r'(\theta)$, and computation load $\frac{r'(\theta)}{K}+\big(1-\frac{r'(\theta)}{K}\big)\lfloor r\rfloor$. For the second group, implement the basic D3C scheme with a  storage space $\lceil r\rceil$, and computation load $\frac{\lceil r\rceil }{K}+\big(1-\frac{\lceil r\rceil}{K}\big)\lceil r\rceil$. %Similar to \eqref{eqn:rtilde}, and \eqref{eqn:rtilde}, the normalized caching and computation can be evaluated to be $r,~ c$ respectively. The communication load $L(c)$ can be evaluated as
%     \begin{IEEEeqnarray}{rCl}
%     &&L(c)\notag\\
%&=&(1-\theta)\frac{1}{\lfloor r\rfloor}\big(1-\frac{r'(\theta)}{K}\big)+\theta\frac{1}{\lceil r\rceil}\big(1-\frac{\lceil r\rceil}{K}\big)\\
%&=&-\frac{1}{\lfloor r\rfloor \lceil r\rceil}\cdot c+\frac{r}{K}\cdot\frac{1}{\lfloor r\rfloor\lceil r\rceil}
%+\big(1-\frac{r}{K}\big)\cdot\bigg(\frac{1}{\lfloor r\rfloor}+\frac{1}{\lceil r\rceil}\bigg), \label{eqn:Lc}
%     \end{IEEEeqnarray}
%where \eqref{eqn:Lc} is obtained by replacing $r'(\theta)$ by $\theta$, and further replacing $\theta$ by $c$ through the connection \eqref{eqn:ctheta}. Notice that,
%     \begin{IEEEeqnarray}{rCl}
%     L(c)|_{c=\frac{r}{K}+(1-\frac{r}{K})\lfloor r\rfloor}&=&\frac{1}{\lfloor r\rfloor}\big(1-\frac{\lfloor r\rfloor}{K}\big),\\
%       L(c) |_{c=c^*(r)}&=&\frac{\lceil r\rceil+\lfloor r\rfloor -r}{\lceil r\rceil \lfloor r\rfloor}-\frac{1}{K}\\
%     &=&L^*(r)
%     \end{IEEEeqnarray}
This results the achievability when $r\notin \mathbb{N}$,    $\lfloor r\rfloor< g\leq g_r$.
\end{enumerate}

Finally, when $g>g_r$, the load $L^*(r)$ can be achieved with the scheme at $g=g_r$.

\section{Conclusion}\label{sec:conclusion}
We proposed a new scheme (the D3C scheme) for distributed computing. It leverages a flexible tradeoff between the three elementary resources \emph{storage space}, \emph{computation load}, and \emph{communication load}. In fact, the D3C scheme achieves the best communication load for any feasible storage space  and computation load, since a matching converse has been established in a following-up work \cite{Yan2018SCC}.

\section*{Acknowledgment}
The work  of Q. Yan and M. Wigger has been supported by the ERC  Grant \emph{CTO Com}.

\ifCLASSOPTIONcaptionsoff
  \newpage
\fi

\end{document}